\documentclass[apj]{emulateapj}

\def\lsim{\mathrel{\lower .85ex\hbox{\rlap{$\sim$}\raise
.95ex\hbox{$<$} }}}
\def\gsim{\mathrel{\lower .80ex\hbox{\rlap{$\sim$}\raise
.90ex\hbox{$>$} }}}
\newbox\grsign \setbox\grsign=\hbox{$>$}
\newdimen\grdimen \grdimen=\ht\grsign
\newbox\laxbox \newbox\gaxbox

\setbox\gaxbox=\hbox{\raise.5ex\hbox{$>$}\llap
     {\lower.5ex\hbox{$\sim$}}}\ht1=\grdimen\dp1=0pt
\setbox\laxbox=\hbox{\raise.5ex\hbox{$<$}\llap
     {\lower.5ex\hbox{$\sim$}}}\ht2=\grdimen\dp2=0pt
\def\gax{\mathrel{\copy\gaxbox}}
\def\lax{\mathrel{\copy\laxbox}}

\shorttitle{GRB 050826 and the Luminosity Distribution of GRBs}
\shortauthors{Mirabal, Halpern, \& O'Brien}

\begin{document}

\title{GRB 050826: A Subluminous Event at
$\MakeLowercase{z} = 0.296$~ Finds its Place in the Luminosity Distribution 
of Gamma-Ray Burst Afterglows}

\author{N. Mirabal\altaffilmark{1}, J. P. Halpern\altaffilmark{1},
\& P. T. O'Brien\altaffilmark{2}}

\altaffiltext{1}{Columbia Astrophysics Laboratory, Columbia University,
New York, NY~10027}
\altaffiltext{2}{Department of Physics and Astronomy, University of Leicester,
Leicester LEI 7RH, UK}

\begin{abstract}
We present the optical identification and spectroscopy of 
the host galaxy of GRB 050826 at a redshift $z = 0.296 \pm 0.001$.  
Image subtraction among observations obtained on three consecutive
nights, reveals a fading object 5 hr after the burst, 
confirming its identification as the optical afterglow of this event.
Deep imaging shows that the optical afterglow is offset
by $0^{\prime\prime}\!.4$ 
(1.76~kpc) from the center of its irregular host galaxy, 
which is typical for long-duration
gamma-ray bursts. 
Combining these results with X-ray measurements acquired by the 
{\it Swift\/} XRT
instrument, we find that GRB 050826 falls entirely within
the subluminous, subenergetic group of long gamma-ray bursts at low 
redshift ($z \lax 0.3$). The results are discussed in the context
of models that possibly account for this trend, including 
the nature of the central engine, the evolution of 
progenitor properties as a function of redshift,
and incompleteness in current 
gamma-ray burst samples.
\end{abstract}


\section{Introduction}

Understanding the progenitor
responsible for gamma-ray bursts
(GRBs) is a fundamental problem in stellar evolution models.
Whereas it is now generally accepted that a fraction of GRBs is
 associated with the deaths of massive stars
\citep{galama,hjorth,stanek}, considerable
uncertainty remains as to what the precise nature of the
progenitor system is, including its evolutionary stage.
The range of potential progenitors seems to be restricted
to rapidly rotating, highly-stripped massive stars,
either in isolation \citep{woosley}, or spun up in close 
binary systems \citep{fryer}.
Unfortunately, neither of these possibilities can yet
be definitely excluded \citep{galyam}.

One key to addressing the origin of GRBs lies with the growing
sample of low-redshift ($z \lax 0.3$)
events \citep[e.g.,][]{mirabal}. According to recent observations
subenergetic, subluminous GRBs/supernovae
dominate the local population of GRB events
\citep{cobb,liang,pian,soderberg}. However, with a handful of low-redshift events,
 it remains unclear whether this trend is due
to  unusual progenitor properties \citep{macfadyen},
or an intrinsic difference in the central engine, {\it i.e.}, black hole
versus magnetar \citep{mazzali,soderberg}. We therefore have set out to 
find the tell-tale signatures of low-redshift bursts in {\it
Swift} afterglows, {\it i.e.}, a bright host galaxy in pre- or postburst 
observations, 
the identification of emission lines associated with a low redshift starburst
galaxy, and/or the rise in supernova light.  
Our ultimate goal is to uncover the redshift distribution, 
host galaxy properties,
and metal content of the nearest progenitor systems.

In this Letter we report optical and X-ray observations of
the nearby GRB 050826, which
we localize to an irregular galaxy at $z = 0.296$. 
We begin with a 
description of the observations and the discovery of the optical transient (OT)
using image subtraction. 
We then discuss the properties
of its host galaxy and X-ray afterglow emission that support
a subluminous classification for this event, when compared to 
cosmic GRBs.
Finally, we consider 
the role of image subtraction in completing the census
of low-luminosity GRBs in nearby galaxies and give an outlook on future work.
Throughout this Letter we assume  $H_0 = $71 km~s$^{-1}$~Mpc$^{-1}$,
$\Omega_m = 0.27$, and $\Omega_{\Lambda} = 0.73$,
corresponding to a luminosity distance
$D_{L}$ = 1517 Mpc.

\section{Observations}

\subsection{$\gamma$-Rays and X-Rays}

GRB 050826
was detected with the {\it Swift\/} Burst Alert Telescope (BAT)
on UT 2005 August 26.2626 \citep{mangano}. The BAT light curve consists of
a multiple-peak structure with $t_{90} = 35 \pm 8$~s \citep{markwardt},
measuring $(4.3 \pm 0.7) \times 10^{-7}$ ergs cm$^{-2}$ in the 15--150 keV band.
While the main burst is weak and hard in the BAT energy
range, the duration is consistent with a classical long 
burst \citep{kouveliotou}.

The {\it Swift\/} X-ray Telescope (XRT) collected data on GRB 050826
from 106~s up to 2.45 days after the
BAT trigger. The processed XRT data presented here
have been assembled from a previous analysis of the X-ray emission
for a sample of {\it Swift\/} GRBs \citep{obrien}. Standard processing
of the data was performed using XRTPIPELINE version 0.8.8 that were then
converted into unabsorbed X-ray fluxes. 

\begin{figure}[t]
\centerline{
  \includegraphics[width=0.97\linewidth]{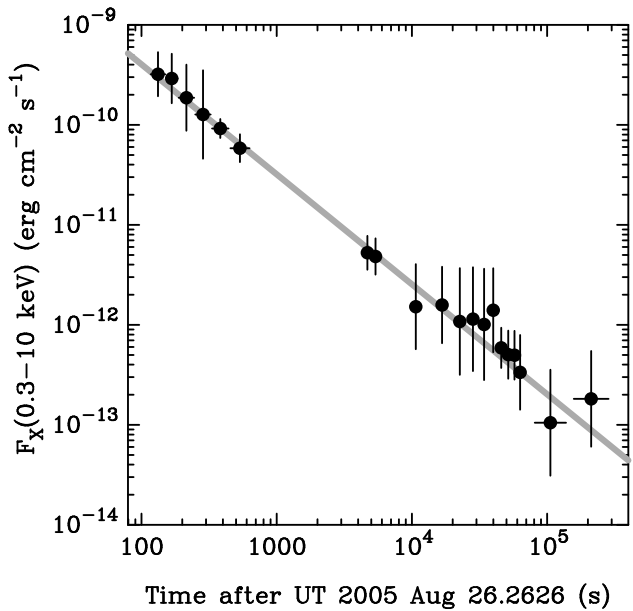}\hfill
}
\caption{XRT light curve (0.3--10 keV) of GRB 050826. The data
are well described by a power-law decay index $\alpha_{\rm X} = -1.10 \pm 0.08$.
}
\label{decay}
\end{figure}

Figure~\ref{decay} shows the resulting XRT light curve in the 0.3--10 keV
bandpass. The temporal decay of the X-ray afterglow is well fitted by a
single power-law model $f \propto t^{\alpha_{X}}$ with a
decay index $\alpha_{X} = -1.10 \pm 0.08$ (see also \citealt*{willing}). From the full spectrum, we
obtain a power-law fit with spectral index $\beta_{X} = -1.27 \pm 0.47$
and $N_{\rm H} = 6.5 \times 10^{21}$ cm$^{-2}$, in excess
of the Galactic value $N_{\rm H} = 2.2 \times 10^{21}$ cm$^{-2}$.

\clearpage
\subsection{Optical}

The {\it Swift\/} UV/Optical Telescope (UVOT)
began observing the field of GRB 050826 just 105 s after the BAT
trigger. No new sources were found within the XRT error circle
to 3-$\sigma$ limiting magnitudes of $B > 21.2$ and
$V > 19.4$  \citep{blustin}. Follow-up optical observations
with the MDM 1.3m telescope commenced on 2005 August 26.450 UT 
and continued for
three consecutive nights until 2005 August 28.480 UT \citep{halpern1}. 
Additional late-time observations of the
region were obtained on 2005 December 25.310 \citep{halpern2} 
and 2007 February 6.135
using the 2.4m and 1.3m MDM telescopes, respectively.

\begin{figure}[t]
\centerline{
  \includegraphics[width=0.93\linewidth]{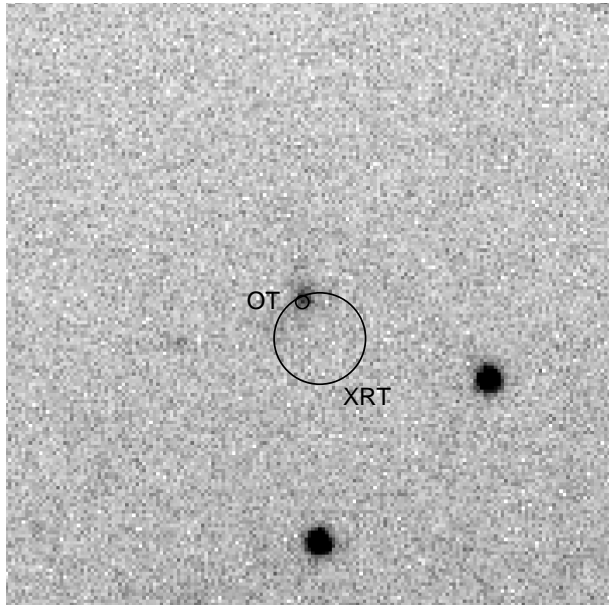}\hfill
}
\caption{$R$-band image of the host galaxy
of GRB 050826 observed with the
MDM 2.4m telescope on UT 2005 December 25.31. The magnitude of
the host is measured to be $R_{\rm host} = 19.67 \pm 0.05$. The localization
of the OT using image subtraction is shown by the inner circle.
Also shown is the final
XRT error position with a $3^{\prime\prime}\!.4$ radius from
\citet{moretti}. The field is $45^{\prime\prime}$ across.}
\label{optical}
\end{figure}

An object not visible on the
Digital Sky Survey is detected  
at $\alpha(\rm J2000) = 05^{h} 51^{m} 01^{s}.58$,
$\delta(\rm J2000) = -02^{\circ} 
38^{\prime} 35^{\prime\prime}\!.8$ on the August 26.472 image. This 
position was originally
$8^{\prime\prime}\!$ away from the initial XRT localization \citep{mangano2}. 
Subsequently, the XRT position (Fig.~\ref{optical}) was
revised to include the optical candidate
within the XRT error circle \citep{moretti,butler}.
To search for optical variability among our images, we performed
image subtraction between the August 26.472 and 28.480 pointings. 
The resulting difference reveals a point-like OT
5 hr after the burst, and shows that the galaxy begins to dominate
 after the August 26.472 epoch.
In Figure~\ref{optical}, we show the OT position derived from
image subtraction overlaid over 
the presumed host galaxy.
A summary of the
optical photometry measured on the residual images is given
in Table~\ref{table1}.

Spectra of the host galaxy were obtained on 
2006 December 24 UT using the Boller and Chivens
CCD Spectrograph (CCDS) mounted on the MDM 2.4m telescope.
A total of three 3600~s exposures were acquired in a $1^{\prime\prime}\!.5$. slit by blind
offset from a nearby field star. The spectra were  processed using
standard procedures in IRAF
\footnote{IRAF is distributed by the National Optical Astronomy Observatories,
    which are operated by the Association of Universities for Research
   in Astronomy, Inc., under cooperative agreement with the National Science Foundation.} and applying the
wavelength calibration from
Xe lamp spectra. Flux calibration was performed using
the spectrophotometric standard Feige~34. Finally, the data were 
dereddened from significant Galactic extinction in this direction, 
$E(B - V) = 0.59$ \citep{schlegel}.
Figure~\ref{spectrum}
shows the summed wavelength-calibrated spectrum of the host
galaxy.

\begin{deluxetable}{lcc}
\tablecolumns{3}
\tablewidth{0pc}
\tablecaption{Optical Photometry of GRB 050826}
\tablehead{
\colhead{Date (UT)}  & \colhead{Filter} & \colhead{Magnitude\tablenotemark{a}} 
}
\startdata
August 26.472 & $R$ & $20.66 \pm 0.15$\\
August 27.473 & $R$ & $>$ 21.24\\
August 28.480 & $R$ & $>$ 21.24\\
\enddata
\tablenotetext{a}{The data have been corrected for Galactic extinction
$A_{R} = 1.57$. 
No extinction intrinsic to the GRB host is included.}
\label{table1}
\end{deluxetable}

\section{Results}

\begin{figure}[t]
\centerline{
  \includegraphics[width=0.97\linewidth]{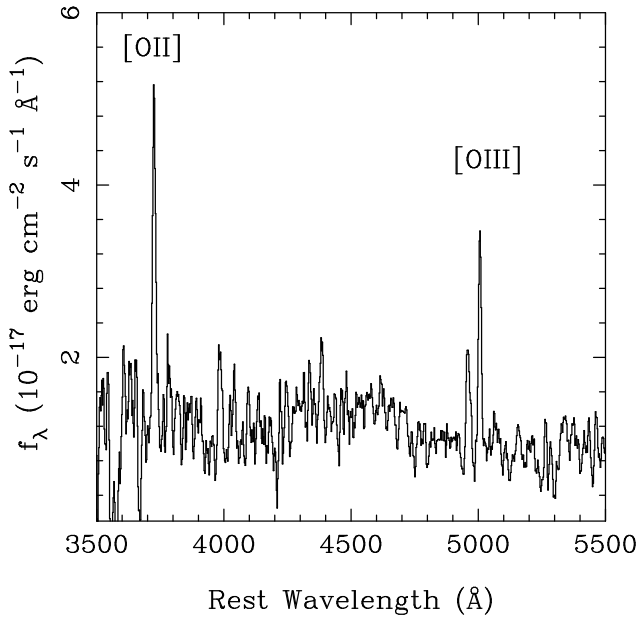}\hfill
}
\caption{Optical spectrum of the host galaxy of GRB 050826 obtained
at the MDM 2.4m telescope on 2006
December 24 UT. Narrow emission lines corresponding to [\ion{O}{2}],
and [\ion{O}{3}] are clearly detected.
The spectrum is corrected for Galactic extinction following a \citet{cardelli}
law. No extinction intrinsic to the GRB host is included.}
\label{spectrum}
\end{figure}

Narrow emission lines
corresponding to [\ion{O}{2}]$\lambda 3727$, and [\ion{O}{3}]$\lambda\lambda 4959,5007$ are seen in the summed
spectrum (Fig.~\ref{spectrum}).  The line strengths are
 similar to those of well-studied GRB
hosts \citep{wiersema}. The weighted mean heliocentric redshift is
 $z = 0.296 \pm 0.001$, thus confirming the initial
redshift interpretation by
\citet{halpern3}. 
Unfortunately, abundance measurements 
require the H$\beta$ intensity, which was impeded by the bright [\ion{O}{1}]
night-sky line at $\approx 6300$ \AA.

At a redshift of $z = 0.296$, the BAT 
$\gamma$-ray fluence $(4.3 \pm 0.7) \times 10^{-7}$ ergs cm$^{-2}$ in the 
15--150 keV band \citep{markwardt} yields an isotropic energy of 
$E_{\rm iso} = (9.1 \pm 1.3) \times 10^{49}$ ergs. 
The simplest afterglow emission model 
consistent with the X-ray observations corresponds to the regime when 
 $\nu_{X} > \nu_{\rm c}$, so that $\beta_{X} = -p/2$ and 
$\alpha_{X} = (2 - 3p)/4$ \citep{granot}. 
Here $p$ is the electron spectral index, and $\nu_{\rm c}$
is the synchrotron cooling frequency. This implies $p = 2.13 \pm 0.1$ 
with either a constant density or a stellar wind circumburst environment. 
We note that a 
burst seen off-axis should show a rising light curve \citep{granot2}, 
which is not detected in this case. 

The lack of a break in the X-ray light curve prior to 2.45 days
postburst, constrains the half-opening angle of the expanding 
jet to $\theta_{0} \gax
0.38~n_{0}^{1/8}$ \citep{sari}, where $n_{0}$ is the circumburst density in 
cm$^{-3}$. Such a wide opening angle appears to strain the degree of 
collimation in the GRB outflow when compared to well-studied events 
\citep{zeh}, however it is difficult to ascertain 
the implications of our results for GRB jet models 
without additional late X-ray data. 
As a result the $\gamma$-ray release in the 15--150 keV 
band is  bracketed by $E_{\gamma}$ = (0.6 -- 9.1) $\times 10^{49}$
ergs. Similarly, the available limits on the afterglow 
luminosity in the 2--10 keV band at $t = 10$ hr (source frame)
correspond to  
$L_{X,10}$ = (0.3 -- 4.6) $\times 10^{43}$ ergs s$^{-1}$.

Inspection of the host galaxy of GRB 050826 in the late-time observations 
reveals a bright core 
and an irregular morphology extended south-east 
(Fig.~\ref{optical}). Photometry of the
host galaxy performed in a $3^{\prime\prime}$ radius aperture
centered on the host nucleus yields 
$R = 21.24 \pm 0.05$ and $V = 22.53 \pm 0.06$, respectively. 
Correcting for the amount of Galactic
extinction, we adopt $R_{\rm host} = 19.67 \pm 0.05$ and  
$V_{\rm host} = 20.59 \pm 0.06$,
as the unextincted magnitudes of the host galaxy. 
Within the current concordance cosmology, 
the implied rest-frame absolute magnitude
corresponds to $M_{B} \approx -19.7$, which is well within the distribution
of GRB host magnitudes 
at redshift $z < 1.2$ \citep{fruchter}. We therefore conclude that the 
host 
luminosity is $L_{\rm host} \approx 0.3 L_{*}$, with
$M_{*} = -21.0$ \citep{christensen}.

From the observed flux in the [\ion{O}{2}] $\lambda 3727$ line
$F_{3727}$ = $(1.1 \pm 0.1) \times 10^{-15}$ ergs cm$^{-2}$ s$^{-1}$,
we derive the line luminosity
$L_{3727}$ = $(2.3 \pm 0.3) \times 10^{41}$ ergs s$^{-1}$.
Following the conversion from \citet{kennicutt}, 
the implied star formation rate corresponds to SFR $\approx (3.2 \pm 1.5)$
$M_{\odot}$ yr$^{-1}$.
Thus, the inferred SFR of the host galaxy
lies in the range 
0.7 -- 12
$M_{\odot}$ yr$^{-1}$  calculated for GRB 
hosts at higher redshifts \citep{christensen}.

The small projected OT displacement from the host  
center $0^{\prime\prime}\!.4$ ($\approx 1.76$~kpc, Figure~\ref{optical}) 
implies that the GRB position correlates with the light of its host 
\citep{bloom}. The chance superposition between the optical
transient and a foreground galaxy of equal or greater brightness 
within the observed offset  
is $\approx 4 \times 10^{-5}$ \citep{huang}, which strengthens its
association with this nearby galaxy. 
As such, 
the host displays an irregular morphology analogous to those observed
in other GRB hosts \citep{fruchter}. Unfortunately, it is difficult to 
determine cleanly whether the south-east extension  
corresponds to neighboring galaxies, or it is related to a 
continuation of the host stellar field.

\begin{figure}[t]
\centerline{
  \includegraphics[width=0.99\linewidth]{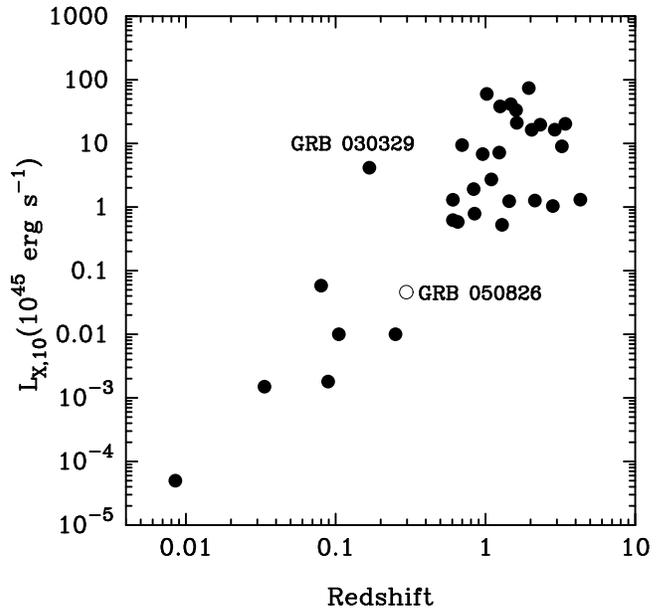}\hfill
}
\caption{Isotropic X-ray luminosity
$L_{X,10}$ in the 2--10 keV band estimated at $t = 10$ hr (source frame) as
a function of redshift ({\it filled circles}) culled from the samples by
\citet{berger}, \citet{nousek}, and \citet{amati}. The  open circle
indicates
the location of GRB 050826 in the distribution. This is a flux-limited sample.
}
\label{luminosity}
\end{figure}

\section{Discussion}

A recent inventory of the prompt and afterglow emission 
of the GRB population reveals that subluminous, subenergetic
GRBs dominate the local population ($z \lax 0.3$) 
of GRB events \citep{cobb,liang,pian,soderberg,kaneko}. 
In order to place GRB 050826 
in the emerging taxonomy of GRBs, we plot its isotropic X-ray luminosity
$L_{X,10}$ in the 2--10 keV band estimated at $t = 10$ hr (source frame) as
a function of redshift 
(Fig.~\ref{luminosity}). For comparison we also show the luminosity 
distribution
of $L_{X,10}$ measurements from the samples amassed by \citet{berger}, 
\citet{nousek}, and \citet{amati}. 

From the collection, it is apparent that GRB 050826
falls below the least luminous GRB at $z \gax 0.3$.
Moreover, for all but one low-redshift ($z \lax 0.3$) burst, the isotropic
afterglow luminosity is bounded by $L_{X,10} \lax 10^{44}$
erg s$^{-1}$. The single exception is GRB 030329, whose  
true luminosity reduces to 
$L_{X,true} \lax 4 \times 10^{43}$ erg s$^{-1}$ 
after the beaming fraction is included \citep{gorosabel}. 
We note that the true X-ray luminosity for 
higher-redshift ($z \gax 0.3$) events will be 
equally dependent on collimation corrections. However, 
collimation-corrected luminosities inferred for $z \gax 0.3$ events are 
consistent with $L_{X,true} \gax 10^{44}$ \citep{berger}.
Thus, on average, 
subluminous GRBs appear to be more prevalent in the local Universe.

Even though we cannot yet pinpoint the origin of this population, it is 
becoming apparent that subluminous GRBs
must be physically different or extreme in properties 
relative to well-studied GRBs at higher redshifts ($z \gax 0.3$). 
One explanation is that there is
an alternative physical channel of stellar collapse 
that leads to subluminous bursts
\citep[e.g.,][]{mazzali,soderberg}. 
The collapsing massive star might, for example,
form a highly-magnetized 
neutron star \citep{usov,thompson} 
rather than a black hole \citep{macfadyen}.  
The greatest obstacle to proving alternative collapse channels for
GRB production is the lack 
of observational signatures that would expose 
the central engine directly during the collapse.

A second possibility is that
progenitor metallicity is what distinguishes 
subluminous events from their high-redshift counterparts \citep{woosley2}. 
At first glance, the sample 
presented in Figure~\ref{luminosity} would seem to point in 
such direction, since
subluminous events should be more prominent when the metallicity is higher,
for example, at lower redshifts \citep{kewley}. 
As it turns out, however, there is little evidence supporting 
the evolution of progenitor properties as a function of redshift

Perhaps the most obvious weakness lies with the incompleteness of 
the current GRB sample. 
It is worth stressing that the redshift trend in
Figure~\ref{luminosity} does not sample low-luminosity events at higher
redshifts \citep{pian,soderberg} and hence the current burst detection
rate might bias the sample toward the more luminous events.
Further we note that 
although the handful of low-redshift events appear to indicate a paucity 
of luminous GRBs in the local Universe, their non-detection does not prove
their demise with the current {\it Swift} detection rate of
one subluminous burst per year (see \S 5).

Additional complications 
arise from contradictory evidence regarding
the metallicity of GRBs and their surroundings. For instance, a number of  
studies suggest a possible correlation between
subluminous GRBs and low-metallicity hosts
\citep{modjaz}. In contrast, abundance estimates from afterglow spectra at
$z \gax 1.5$  allow the interstellar medium (ISM) 
surrounding the GRB event to reach solar metallicity \citep{prochaska}.
One caveat is that the rotational energy budget prior to the GRB onset
may be ultimately controlled by the iron abundance of the progenitor   
\citep{vink}. Unfortunately, to the best of our knowledge,  
there are no conclusive identifications of iron or any other metal
lines forged by the GRB progenitor
\citep{mirabal2,sako}. We conclude that 
at least two alternatives for subluminous burst production 
are broadly consistent with current measurements. As a result, the origin
of subluminous GRBs remains unsettled. 

\section{Conclusions and Future Work}

The
optical and X-ray observations of GRB 050826 we have presented 
confirm a general trend 
in which subluminous explosions dominate the local population ($z \lax 0.3$)
 of long-duration GRB events. 
Optical imaging reveals that the OT associated with GRB 050826 is 
located within an irregular,
star-forming 
host galaxy with a rest-frame $B$-band luminosity 
$L_{\rm host} \approx 0.3 L_{*}$.    
Together, these findings make the host galaxy of GRB 050826  
an excellent target for   
high-resolution spectral studies 
at the site of the explosion. 

In the quest to understand the origin of subluminous GRBs, 
it appears crucial to optimize future search strategies 
of subluminous GRBs at higher redshifts
($z \ga 0.3$). In parallel, it would be prudent to 
explore numerically various afterglow observables as a function 
of accretion rate, and  energy output from the central engine. 
This may lead to a better understanding of the link between the
central engine and the afterglow luminosity distribution. 

Lastly, a more complete analysis is still limited by the reduced number 
of low-redshift GRBs observed to date. We expect  
image subtraction techniques will play an important role in completing
the census of subluminous GRBs in nearby galaxies.
In particular, observations with future synoptic telescopes 
such as Pan-Starrs \citep{kaiser} and the Large Synoptic Survey Telescope 
\citep{tyson} have the potential of detecting 
additional nearby bursts that might have been missed with the 
localization rate of GRB missions. Assuming a rate of subluminous
events   
$230^{+490}_{-190}$ Gpc$^{-3}$ yr$^{-1}$ \citep{soderberg}, a telescope cadence
that covers a large portion of the available sky every three nights, and a 
limiting magnitude $V_{\rm lim} = 23.5$, shows that a dedicated
synoptic telescope could discover $2553^{+5439}_{-2190}$ 
events as bright as GRB 060218/SN 2006aj \citep{mirabal}
per year  
out to a maximum distance $D_{\rm max} = 3$~Gpc.
Our results therefore suggest that current GRB/supernova rates could be enhanced
by at least 2 orders of magnitude, should these exist. 
Such an improvement is likely 
to reshape GRB/supernova research dramatically.

\acknowledgments
This work was supported by  
{\it Swift\/} grant NNH05ZDA001N.

\end{document}